\documentclass[hyper,aps,eqsecnum,showpacs,superscriptaddress,12pt]{revtex4}
\usepackage{mathrsfs}
\usepackage{amsmath,amssymb,bm,comment}
\usepackage{hyperref}
\usepackage{graphicx}
\usepackage{dcolumn}
\usepackage{bm}
\usepackage{slashed}
\usepackage[a4paper,ignoreall]{geometry}
\usepackage[usenames,dvipsnames]{xcolor}
\usepackage{color}
\usepackage{subfigure}
\usepackage{ulem}

\begin{document}
\title{Nuclear dependencies of azimuthal asymmetries in the Drell-Yan process}

\author{Long Chen}
\affiliation{School of Physics \& Key Laboratory of Particle Physics and Particle Irridiation (MOE),
Shandong University, Jinan, Shandong 250100, China}
\author{Jian-hua Gao}
\email{gaojh@sdu.edu.cn}
 \affiliation{Shandong Provincial Key Laboratory of Optical Astronomy and Solar-Terrestrial Environment, School of Space Science and Physics, Shandong University at Weihai, Weihai 264209, China}
 \affiliation{Key Laboratory of Quark and Lepton Physics (MOE), Central China Normal University,
Wuhan 430079, China}
\author{Zuo-tang Liang}
\affiliation{School of Physics \& Key Laboratory of Particle Physics and Particle Irridiation (MOE), Shandong University, Jinan, Shandong 250100, China}

\date{\today}

\begin{abstract}
We study nuclear dependencies of azimuthal asymmetries in the Drell-Yan lepton pair production in
nucleon-nucleus collisions with polarized nucleons.
We use the ``maximal two-gluon correlation approximation,"
so that we can relate the transverse-momentum-dependent quark distribution in a nucleus  to that
in a nucleon by a convolution with a Gaussian broadening.
We use the Gaussian ansatz for the transverse momentum dependence of such quark distribution functions
and obtain the numerical results for the nuclear dependencies.
These results show that the {$q_T$-integrated} azimuthal asymmetries are suppressed.
\end{abstract}

\maketitle
\section{introduction}

Both the deep inelastic scattering (DIS) off hadrons and the Drell-Yan (DY) process in hadron-hadron collision have been
playing very important roles in studying the structure of hadrons and the dynamics of quantum chromodynamics (QCD).
Correspondingly, the DIS off nuclei and the DY process in hadron-nucleus collisions are also very important in studying
the nuclear structure and the properties of  cold nuclear matter.
By studying the corresponding semi-inclusive processes,
we can study not only the longitudinal but also the transverse momentum dependence of the parton distribution functions.
In this connection, the semi-inclusive DY process is even more suitable to study
the structure of hadrons or that of nuclei because no fragmentation function is involved.
Azimuthal asymmetries are often sensitive physical variables for such studies and
thus have attracted much attention~\cite{Georgi:1977tv,Cahn:1978se,Berger:1979kz,Oganesian:1997jq,Chay:1997qy,Collins:1977iv,Collins:1978yt,Lam:1978pu,
Lam:1978zr,Lam:1980uc,Fries:1999jj,Gelis:2006hy,Liang:2006wp,Zhou:2009jm}.

When a parton transmits through  nuclear matter, the multiple gluon scattering with the nuclear matter
leads to  energy loss and transverse momentum broadening~\cite{Bodwin:1988fs,Luo:1992fz,Baier:1996sk,Guo:1998rd,Wiedemann:2000za,
Guo:2000nz,Wang:2001ifa,Fries:2002mu,Majumder:2007hx,Liang:2008vz,D'Eramo:2010ak,D'Eramo:2011zz,D'Eramo:2011zzb}.
The multiple parton interaction results in also nuclear dependencies of azimuthal asymmetries.
This provides a good alternative probe of properties of the nuclear matter.
The nuclear dependence of the azimuthal asymmetry in semi-inclusive deep inelastic scattering (SIDIS) has been studied
recently~\cite{Gao:2010mj,Song:2010pf, Gao:2011mf}.
In this paper, we extend the study in Refs.\cite{Gao:2010mj,Song:2010pf} to the DY process in nucleon-nucleus collisions.
In Sec. II, we review  the result of the differential cross section in the DY process with the polarized nucleon beam
in terms of the transverse-momentum-dependent(TMD) parton distributions up to twist-2 level.
In Sec. III, we study the  nuclear dependence of the angular distribution of the DY lepton pair by relating
the TMD quark distributions in a nucleus  to that in a nucleon. We also illustrate the numerical results with
an ansatz of the TMD parton  distributions in a Gaussian form.  We give a brief summary in Sec. IV.


\section{Differential cross section and azimuthal asymmetries}

We consider the semi-inclusive DY process in nucleon-nucleus collisions with the transversely or longitudinally polarized nucleon beam,
\begin{eqnarray}
N(p_1,s)+A(A p_2) &\rightarrow& \gamma^*(q)+X
\rightarrow l^+(l)+l^-(l')+X,
\end{eqnarray}
where $p_1$, $p_2$, $q$, $l$, and $l'$ are the four-momenta of the beam nucleon,
one nucleon in the nucleus target, the virtual photon, the anti-lepton, and the lepton, respectively,
and $s$ denotes the polarization  vector of the incident nucleon.
We use the light cone coordinate by introducing two light like vectors, $n_{+}=[1,0,\vec{0}_{\perp}]$ and $ n_{-}=[0,1,\vec{0}_{\perp}]$,
and express the momenta $p_1$ and $p_2$ as
\begin{eqnarray}
p_1^{\mu}&=&p_1^+ n_{+}^{\mu} +\frac{M^2}{2p_1^+}n_{-}^{\mu},\\
p_2^{\mu}&=&\frac{M^2}{2p_2^-}n_{+}^{\mu}+p_2^-n_{-}^{\mu},
\end{eqnarray}
where $p^+ = p\cdot n_{-}$, $p^- = p\cdot n_{+}$, and $M$ denotes the mass of the nucleon.
We restrict our study to the kinematic region where the transverse momentum ${q}_T $ of
the DY pair is much less than its invariant mass $Q=\sqrt{q^2}$.
In this case, the differential cross section for the semi-inclusive DY process can be calculated in the framework
of the TMD factorization theorem \cite{Collins:1981uk,Ji:2004xq}.
Such calculations have been carried out for hadron-hadron collisions;
the results can be found in, e.g., Refs.\cite{Boer:1999mm,Arnold:2008kf,Lu:2011cw}.
We note that such calculations can be extended to nucleon-nucleus collisions in a straightforward way and,
at the twist-2 level, the differential cross section is given by
\begin{eqnarray}
\label{sigma}
%
&&\frac{ d\sigma}{d^2\Omega d^2q_Tdx_1dx_2}
=\frac{\alpha_{em}^2}{4Q^2}\left\{ (1+\cos^2\theta)\mathcal{F}_0\left[f_{1},f_{1}\right]
+\sin^2\theta \cos 2\phi \mathcal{F}_1\left[h_{1}^{\perp},h_{1}^{\perp}\right] \right. \nonumber\\
&&\phantom{XXXX}
+\lambda_s\sin^2\theta \sin2\phi \mathcal{F}_1\left[h_{1L}^{\perp},h_{1}^{\perp}\right]
-|\vec{s}_{T}| (1+\cos^2\theta)\sin\phi_s\mathcal{F}_2\left[f_{1T}^{\perp},f_{1}\right]\nonumber\\
&&\phantom{XXXX}
\left. +|\vec{s}_{T}| \sin^2\theta\sin(2\phi-\phi_s)\mathcal{F}_3\left[h_{1},h_1^{\perp}\right]
+|\vec{s}_{T}| \sin^2\theta \sin(2\phi+\phi_s)\mathcal{F}_4 \left[h_{1T}^{\perp},h_{1}^{\perp}\right] \right\},\ \ \ \ \ \  %
\end{eqnarray}
where $\lambda_s$ and $\vec{s}_{T}$ are, respectively, the helicity
and the transverse polarization vector of the nucleon; $\theta$,
$\phi$, and $\phi_s$ are, respectively, polar and azimuthal angles of
the lepton pair and  azimuthal angle of the polarization vector of the
nucleon with respect to the transverse vector $\vec q_T$ in the
Collins-Soper frame. The $\mathcal{F}_j[f,h]$'s ($j=0$ through $4$)
are functionals of $f(x,\vec k_T)$ and $h(x,\vec k_T)$ that are defined
as convolutions weighted by $\chi_j(\vec q_T,\vec k_{1T},\vec k_{2T})$,
\begin{eqnarray}
\mathcal{F}_j[f,h]
&\equiv&\frac{1}{3}\sum_{a}e_a^2\int \frac{d^2k_{1T}}{(2\pi)^2} \frac{d^2k_{2T}}{(2\pi)^2}
\delta^2(\vec{k}_{1T}+\vec{k}_{2T} - \vec{q}_{T})\chi_j(\vec q_T,\vec{k}_{1T},\vec{k}_{2T})\nonumber\\
& &\times \left[f^N(x_1,\vec{k}_{1T};a) h^A(x_2,\vec{k}_{2T};\bar a)+ f^N(x_1,\vec{k}_{1T};\bar a)  h^A(x_2,\vec{k}_{2T};a)\right],\hspace{0.5cm}
\end{eqnarray}
where $f$ and $h$ are the  TMD distribution and/or correlation functions of quarks or anti-quarks.
The superscript $N$ or $A$ denotes whether it is for the nucleon or the nucleus, and
$a$ and $\bar a$ in the arguments denote the flavor of the quark and whether it is for the quark or the anti-quark.
The weights $\chi_j$ are given by
\begin{eqnarray*}
\chi_0 &=& 1, \\
\chi_1 &=& \frac{1}{M^2}\left[2\left(\vec{k}_{1T}\cdot {\hat{\vec q}}_{T}\right) \left(\vec{k}_{2T}\cdot \hat{\vec{q}}_{T}\right)-\vec{k}_{1T}\cdot \vec{k}_{2T}\right],\\
\chi_2 &=& \frac{1}{M} \vec{k}_{1T}\cdot {\hat{\vec q}}_{T}, \\
\chi_3 &=& \frac{1}{ M} \vec{k}_{2T}\cdot \hat{\vec{q}}_{T}, \\
\chi_4 &=& \frac{1}{2M^3}\left[4\left(\vec{k}_{1T}\cdot {\hat{\vec q}}_{T}\right)^2 \left(\vec{k}_{2T}\cdot \hat{\vec{q}}_{T}\right)
 -2\left(\vec{k}_{1T}\cdot \hat{\vec{q}}_{T}\right)\left(\vec{k}_{2T}\cdot\vec{k}_{1T}\right)-\vec{k}_{1T}^2\left(\vec{k}_{2T}\cdot \hat{\vec{q}}_{T}\right)\right],
\end{eqnarray*}
where $\hat{\vec q}_T\equiv \vec q_T /\sqrt{\vec q_T^{\ 2}}$.
All the parton distribution and correlation functions, $f$'s and $h$'s, given in  Eq.(\ref{sigma}) are defined by
the twist-2 decomposition of the quark correlation matrix \cite{Mulders:1995dh,Goeke:2005hb,Bacchetta:2006tn},
\begin{eqnarray}
\label{phi}
{\Phi}(x,\vec{k}_{T},{s})
&=&\left\{  f_1 \frac{\slashed {n}_+}{2} -f_{1T}^{\perp}
\frac{\epsilon_{T}^{\rho\sigma}k_{T\rho}s_{T\sigma}}
{M} \frac{\slashed {n}_+}{2}+g_{1L}\lambda_{s} \frac{\gamma_5
\slashed{n}_+}{2}-g_{1T}\frac{k_{T}\cdot s_{T}}{M}
\frac{\gamma_5 \slashed{n}_+}{2}
+h_{1T}\frac{\gamma_5\slashed{s}_{T}\slashed{n}_+}{2}\right.\nonumber\\
& & \left.+h_{1L}^{\perp}\lambda_{s}\frac{1}{M}\frac{\gamma_5\slashed{k}_{T}\slashed{n}_+}{2}
-h_{1T}^{\perp}\frac{k_{T}\cdot s_{T}}{M}\frac{1}{M}\frac{\gamma_5
\slashed{k}_{T}\slashed{n}_+}{2}
+h_{1}^{\perp}\frac{1}{M}\frac{i\slashed{k}_{T}\slashed {n}_+}{2} \right\},
\end{eqnarray}
where $ \epsilon _{T}^{\mu\nu}=\epsilon ^{\mu\nu\rho\sigma}n_{+\rho}n_{-\sigma}$ with the total antisymmetric tensor $\epsilon^{0123}=+1$ and  the definition of the components of ${\Phi}(x,\vec{k}_{T},{s})$ are given by,
\begin{eqnarray}
\label{phi}
\Phi_{\alpha\beta}(x,\vec k_T,s) & \equiv& \int \frac{p^+ dy^-}{2\pi} \frac{d^2y_\perp}{(2\pi)^2}
e^{ixp^+y^- -i\vec  k_\perp\cdot \vec y_\perp}
\langle N,s \mid \bar\psi_\beta(0){\cal L}(0,y)\psi_\alpha(y)\mid N,s \rangle, \ \ \ \ \ \ \
\end{eqnarray}
where ${\cal L}(0,y)$ is the gauge link that is necessary to  ensure the gauge invariance of the matrix. In the DY process,
the gauge link  in covariant gauge is given by
\begin{eqnarray}
\label{link}
{\cal L}(0,y)=\mathcal{L}[y^-,\vec{y}_\perp;-\infty,\vec{y}_\perp ]  \mathcal{L}^\dagger[0,\vec{0}_\perp;-\infty,\vec{0}_\perp ],
\end{eqnarray}
where
\begin{equation}
\label{TMDGL}
\mathcal{L}[ y^-,\vec{y}_\perp;-\infty,\vec{y}_\perp]\equiv
P \exp \left(- i g \int^{y^-}_{-\infty} d \xi^{-} A^+ ( \xi^-, \vec{y}_\perp)
\right) \, .
\end{equation}
It should be noted that the distribution function $h_1$ in Eq. (\ref{sigma}) is defined as the mixture of
$h_{1T}$ and $h_{1T}^\perp$,
\begin{eqnarray}
\label{h1}
h_1(x,\vec k_{T}) &\equiv& h_{1T}(x,\vec k_{T})+\frac{\vec k_{T}^2}{2M^2}h_{1T}^\perp(x,\vec k_{T}).
\end{eqnarray}
We see that the differential cross section is determined by six TMD quark and anti-quark distributions and
correlation functions $f_1(x,\vec k_T)$,  $f_{1T}^\perp(x,\vec k_T)$, $h_{1}^\perp(x,\vec k_T)$, $h_{1}(x,\vec k_T)$,
$h_{1T}^\perp(x,\vec k_T)$, and $h_{1L}^\perp(x,\vec k_T)$.
Each of them represents a given aspect of the parton structure of the nucleon, e.g.,
$f_{1T}^\perp(x,\vec k_T)$ is the Sivers function\cite{Sivers:1989cc,Sivers:1990fh} which describes
the correlation between the transverse momentum distribution and the transverse polarization of the nucleon,
and  $h_{1}^\perp(x,\vec k_T)$ is the Boer-Mulders function \cite{Boer:1997nt}, which describes the correlation between the
transverse quark momentum distribution and the transverse quark polarization in an unpolarized nucleon.
{In the TMD factorization formalism, all these distribution functions are unknown and
cannot be calculated perturbatively.
They can usually be obtained from parameterizations of experimental data or from model calculations
(see, e.g., Refs. \cite{Pasquini:2006iv,Anselmino:2007fs,Anselmino:2008jk,Gamberg:2003ey,Gamberg:2007wm,Bacchetta:2007wc,Avakian:2007xa,
Pasquini:2008ax,Bacchetta:2008af,Courtoy:2008vi,Courtoy:2008dn,Courtoy:2009pc,Avakian:2008dz,Anselmino:2008sga,
Arnold:2008ap,Efremov:2009ze,She:2009jq,Jakob:1997wg,Efremov:2003eq,Yuan:2003wk,Pobylitsa:2003ty,Efremov:2004qs,
Pasquini:2010af}).
We should also note that these correlation functions such as the Sivers and the Boer-Mulders functions
reflect not only the intrinsic motion of a parton inside a nucleon but also the multiple gluon scatterings
(referred as initial or final state interaction) contained in the gauge link.
In this connection, we recall the proof of Collins \cite{Collins:1992kk}
that Sivers function is zero if we take the gauge link as unity.
The same conclusion applies to the Boer-Mulders function.
An intuitive but semiclassical picture for the non-zero Sivers function or the existence of left-right
single-spin asymmetry was proposed in the 1990s\cite{BLM93} where one invokes the orbital angular momentum
of quark and differentiates between the ``front" and ``back" surface of a nucleon due to the initial state interaction.
This agrees qualitatively with the field theoretical calculations later on by Brodsky, Hwang, and Schmidt \cite{Brodsky:2002rv},
where they use a non-zero orbital angular momentum of quark and take the initial state interaction into account explicitly
and obtain a nonzero result for the Sivers function.
Apparently, the same picture lead also to nonzero Boer-Mulders function. }

We can integrate over the polar angle $\theta$  in Eq.(\ref{sigma}) and obtain
\begin{eqnarray}
\label{sigma-2}
&&\frac{ d\sigma}{d\phi d^2q_Tdx_1dx_2}
=\frac{\alpha_{em}^2}{3Q^2}\left\{ 2\mathcal{F}_0\left[f_{1},f_{1}\right]
+ \cos 2\phi \mathcal{F}_1\left[h_{1}^{\perp},h_{1}^{\perp}\right] \right. \nonumber\\
&&\phantom{XXXX}
+\lambda_{s} \sin2\phi \mathcal{F}_1\left[h_{1L}^{\perp},h_{1}^{\perp}\right]
-2 |\vec{s}_{T}| \sin\phi_s\mathcal{F}_2\left[f_{1T}^{\perp},f_{1}\right]\nonumber\\
&&\phantom{XXXX}
\left. +|\vec{s}_{T}| \sin(2\phi-\phi_s)\mathcal{F}_3\left[h_{1},h_1^{\perp}\right]
+|\vec{s}_{T}| \sin(2\phi+\phi_s)\mathcal{F}_4 \left[h_{1T}^{\perp},h_{1}^{\perp}\right] \right\}.\ \ \ \ \ \  %
\end{eqnarray}
We see that there are five kinds of different azimuthal asymmetries
that are given by the average of $\cos 2\phi$, $\sin2\phi$, $\sin\phi_s$, $\sin(2\phi-\phi_s)$,
and $\sin(2\phi+\phi_s)$ respectively.
In terms of parton distribution and correlation functions, they are given by
\begin{eqnarray}
\label{AAA}
A_{NA}^{\cos 2\phi} &= & \frac{ \mathcal{F}_1\left[ h_{1}^{\perp}, h_{1}^{\perp}\right]}  {4\mathcal{F}_0\left[f_{1},f_{1}\right]},\\
A_{NA}^{\sin 2\phi} &=& \lambda_s \frac{ \mathcal{F}_1\left[  h_{1L}^{\perp}, h_{1}^{\perp}\right]} {4\mathcal{F}_0\left[f_{1},f_{1}\right]},\\
A_{NA}^{\sin\phi_s} &=& -|\vec{s}_{T}|\frac{\mathcal{F}_2\left[ f_{1T}^{\perp},f_{1}\right]} {2\mathcal{F}_0\left[f_{1}, f_{1}\right]}, \\
A_{NA}^{\sin(2\phi-\phi_s)} &=& |\vec{s}_{T}|\frac{ \mathcal{F}_3\left[h_{1}, h_{1}^{\perp}\right]} {4\mathcal{F}_0\left[f_{1}, f_{1}\right]},\\
A_{NA}^{\sin(2\phi+\phi_s)} &=& |\vec{s}_{T}|\frac{\mathcal{F}_4\left[ h_{1T}^{\perp}, h_{1}^{\perp}\right]} {4\mathcal{F}_0\left[f_{1},f_{1}\right]},
\end{eqnarray}
where the subscript $NA$ denotes the nucleon-nucleus collisions.
They differ from those for nucleon-nucleon collisions only by the quark distribution and/or correlation functions as  given by Eq. (\ref{phi}).

We also note that these asymmetries exist in collisions with the unpolarized,  longitudinally polarized, and transversely polarized nucleon beams.
In the unpolarized case, only $\cos 2\phi$ exists and is determined by the Boer-Mulders functions $h_{1}^\perp$.
There is one single spin-asymmetry (SSA) $\sin 2\phi$ in collisions with the longitudinally polarized beam,
and it is determined by the longitudinal transversity $h_{1L}^{\perp}$ and the Boer-Mulders function $h_{1}^\perp$.
There are three SSAs in collisions with the transversely polarized beam.
They are represented by $\sin\phi_s$, $\sin(2\phi-\phi_s)$, and $\sin(2\phi+\phi_s)$.
The well-known SSA $\sin\phi_s$ is determined by the Sivers function $f_{1T}^{\perp}$,
while $\sin(2\phi+\phi_s)$ and $\sin(2\phi-\phi_s)$ are determined by the Boer-Mulders function $h_{1}^\perp$
together with the pretzelosity $h_{1T}^\perp$ or $h_{1}$ mixed from $h_{1T}$ and $h_{1T}^\perp$ respectively.
Although we still do not know much about them, these functions have been studied in semi-inclusive DIS \cite{Mkrtchyan:2007sr,Airapetian:2009ae,Avakian:2010ae,Alekseev:2010rw} and
some rough parameterizations have already been made
\cite{Collins:2005rq,Vogelsang:2005cs,Anselmino:2008sga,Anselmino:2007fs,Anselmino:2008jk}.

\section{Nuclear dependence}

It has been shown  \cite{Liang:2008vz,Gao:2010mj}  that multiple gluon scattering represented by the gauge link given by Eq. (\ref{link})
leads to a strong nuclear dependence of the TMD parton distribution and/or correlation functions.
Because all the asymmetries presented above are functionals of these parton correlation functions, we expect strong
nuclear dependence of these asymmetries.
We discuss them in the following.

\subsection{Nuclear dependence of the TMD parton correlation function}

In Ref. \cite{Liang:2008vz}, with the assumption  that the nucleus is large and weakly bound,
the multiple-nucleon correlation can be neglected; the nuclear effect can only arise from the final state
interaction in the form of multiple gluon scattering that is encoded into the gauge link in the definition
of the TMD parton distributions.
The important trick for the derivations is that the  TMD  quark distributions in nucleons or nuclei can be rewritten
as a sum of higher-twist collinear parton matrix elements,
\begin{eqnarray}
f_1^A(x, \vec k_T)&=&\int \frac{dy^-}{2\pi}e^{ixp^+y^-}
\langle A \mid \bar\psi(0)\frac{\gamma^+}{2}
e^{\vec W_T(y^-) \cdot\vec\nabla_{k_T}}
\psi(y^-)\mid A \rangle
\delta^{(2)}(\vec k_T),
\end{eqnarray}
where $\vec W_T(y^-)$ is the parton transport operator and is given by
\begin{eqnarray}
\vec W_T(y^-)&\equiv& i\vec D_T(y^-)+g\int_{-\infty}^{y^-}d\xi^- \vec F_{+T}(\xi^-),
\label{eq:transop}
\end{eqnarray}
with $\vec D_T(y^-)$ being the covariant derivative. For simplicity, we have chosen
the light-cone gauge in which the collinear gauge link  disappears in the above.
The nuclear effect arises when the the parton transport operator acts on the different nucleons.
Under the ``maximum two-gluon correlation approximation" \cite{Liang:2008vz},  the nuclear TMD parton distribution
has been expressed in terms of a Gaussian convolution of the same TMD distribution in a nucleon, i.e.,
\begin{eqnarray}
\label{relation1}
 f_{1}^A(x,\vec{k}_{T})&\approx &\frac{A}{\pi \Delta_{2F}}\int d^2l_{T}
e^{\frac{-(\vec{k}_{T}-\vec{l}_{T})^2}{\Delta_{2F}}}  f_{1}^N(x,\vec{l}_{T}),
\end{eqnarray}
where  $\Delta_{2F}$ denotes  the total average squared transverse momentum broadening.
Furthermore, it has been shown in Ref.\cite{Gao:2010mj} that the relation can be extended to a much more
general case so that
\begin{eqnarray}
\label{phiAN}
\Phi^{A}(x,\vec k_T)
& \approx & A \exp\left[\frac{\Delta_{2F}}{4}\nabla_{ k_T}^2\right] \Phi^{N}(x,\vec k_T)\nonumber\\
&=&\frac{A}{\pi \Delta_{2F}}
\int d^2l_T e^{-(\vec k_T -\vec l_T)^2/\Delta_{2F}}\Phi^{N}(x,\vec l_T),
\label{tmdgeneral}
\end{eqnarray}
where the components of the matrix $\Phi^A$ are defined in Eq.(\ref{phi}).
{A somewhat different derivation can be found in Ref. \cite{Majumder:2007hx,Majumder:2007ne}
where they resumed all the possible gluon exchange attached to different nucleons
in the nucleus and obtained a diffusion equation that leads to the Gaussian convolution,
the same result as that obtained in Ref. \cite{Liang:2008vz}.}
Recently it has been shown \cite{Schafer:2013mza}
that such simple Gaussian convolution will be broken by the process-dependent gauge links
in cold nuclear matter when the finite volume effects are considered.
{In Ref. \cite{Liang:2008vz} and the current paper,
we consider the limiting case of very large nuclei and neglect the finite volume effects.
In the approach presented in Ref. \cite{Schafer:2013mza},
the authors try to study the finite volume effects
where they have to take some specified model and consider the process-dependent gauge links.
Their results are much more complicated than those obtained in Refs. \cite{Liang:2008vz} and \cite{Majumder:2007hx,Majumder:2007ne}
and they find that the simple Gaussian convolution is broken.
Here, in this paper, we consider the simple case as considered in Refs. \cite{Liang:2008vz} and \cite{Majumder:2007hx,Majumder:2007ne}
and take the simple Gaussian convolution in the following.}

For the DY azimuthal asymmetries presented in last section for nucleon-nucleus collisions,
besides the TMD quark distribution $f_{1}^{A}$,    the Boer-Mulders distribution $h_{1}^{A\perp}$
in the nucleus is also involved.
From the decomposition in  Eq. (\ref{phi}), we can express the Boer-Mulders distribution as the following
\begin{eqnarray}
h_{1}^{N/A\perp}(x,\vec{k}_{T})&=&\frac{M^2}{2k_T^2}\textrm{Tr}\left[{i\slashed{k}_{T}\slashed {n}_- \Phi^{N/A}} \right].
\end{eqnarray}
From the relation (\ref{phiAN}), we can show that the Boer-Mulders function in the nucleus  is related to that in the nucleon
in the exactly same way as the twist-3 distribution in Ref. \cite{Gao:2010mj}, i.e.,
\begin{eqnarray}
\label{relation2q}
h_{1}^{A\perp}(x,\vec{k}_{T})&=&\frac{A}{\pi \Delta_{2F}}
\int d^2 l_T \frac{(\vec k_T\cdot\vec l_T)}{\vec k_T^{2}}
e^{-(\vec k_T -\vec l_T)^2/\Delta_{2F}}h_{1}^{N\perp}(x,\vec l_T).
\end{eqnarray}
If we take the Gaussian ansatz for the transverse momentum dependence, i.e.,
\begin{eqnarray}
 f_{1}^N(x,\vec{k}_{T})&=&\frac{1}{\pi\alpha}f_{1}^N(x)
e^{-\frac{\vec k_{T}^2}{\alpha}},\\
h_{1}^{N\perp}(x,\vec{k}_{T} )&=&\frac{1}{\pi\beta}h_{1}^{N\perp}(x)
e^{-\frac{\vec k_{T}^2}{\beta}},
\end{eqnarray}
where we have assumed different flavors have the same Gaussian widths  for the same types of  TMD distributions and we have suppressed the flavor index.
We obtain from Eqs. (\ref{relation1}) and (\ref{relation2q}) that
\begin{eqnarray}
 f_{1}^{{A}}(x,\vec{k}_{T})
&=&\frac{A}{\pi(\alpha +\Delta_{2F})}f_{1}^N(x)e^{-\frac{\vec k_{T}^2}{\alpha
+\Delta_{2F}}},\\
h_{1}^{{{A}}\perp}(x,\vec{k}_{T})
&=&\frac{A\beta}{\pi(\beta +\Delta_{2F})^2}h_{1}^{N\perp}(x)
e^{-\frac{\vec k_{T}^2}{\beta+\Delta_{2F}}}.
\end{eqnarray}

\subsection{Nuclear dependence of the azimuthal asymmetry}

It follows that azimuthal asymmetry $\cos 2\phi$ in nucleon-nucleon and
nucleon-nucleus collisions are  given by, respectively,
\begin{eqnarray}
\label{ANNcos2phi}
A_{NN}^{\cos2\phi}&=& \frac{1}{4M^2}\frac{\alpha}{4\beta}
\frac{\mathcal{S}\left[h_{1}^{\perp},{h}_{1}^{\perp}\right]}{\mathcal{S}\left[f_{1},{f}_{1}\right]}\vec{q}_{T}^{\ 2}
e^{-\frac{\alpha-\beta}{2\alpha\beta}\vec{q}_{T}^{\ 2}},\\
\label{ANAcos2phi}
A_{NA}^{\cos2\phi}&=&\frac{1}{4M^2}\frac{\beta^2(2\alpha+\Delta_{2F})}
{(2\beta+\Delta_{2F})^3}
\frac{\mathcal{S} \left[h_{1}^{\perp},{h}_{1}^{\perp}\right]}{\mathcal{S}\left[f_{1},{f}_{1}\right]}\vec{q}_{T}^{\ 2}
e^{-\frac{2(\alpha-\beta)}{(2\alpha+\Delta_{2F})
(2\beta+\Delta_{2F})}\vec{q}_{T}^{\ 2}},\hspace{1cm}
\end{eqnarray}
where we have defined a shorthand notation,
\begin{eqnarray}
\mathcal{S}\left[f_1,f_1 \right]
&\equiv&\frac{1}{3}\sum_{a}e_a^2
\left[f_1^N(x_1;a)  f_1^N(x_2;\bar a)+f_1^N(x_1;\bar a)  f_1^N(x_2; a)\right].\hspace{0.5cm}
\end{eqnarray}
It is obvious that we can obtain $A_{NN}^{\cos2\phi}$ by simply setting $\Delta_{2F}=0$ in $A_{NA}^{\cos2\phi}$. Hence
we only present $A_{NA}$ in the other azimuthal asymmetries in the following discussion.

\begin{figure}
\subfigure[]{\includegraphics[width=0.6\textwidth]{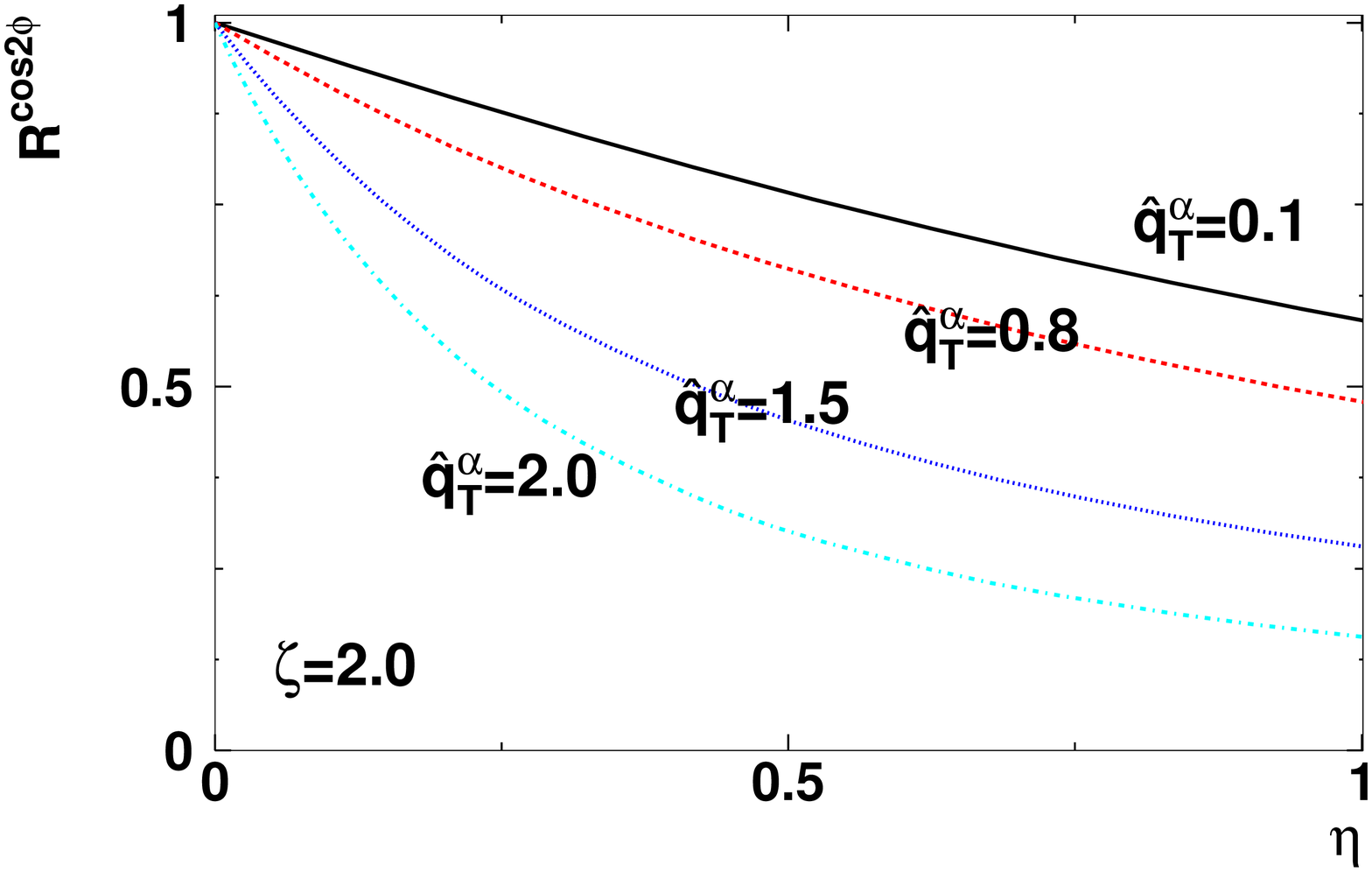}}
\subfigure[]{\includegraphics[width=0.6\textwidth]{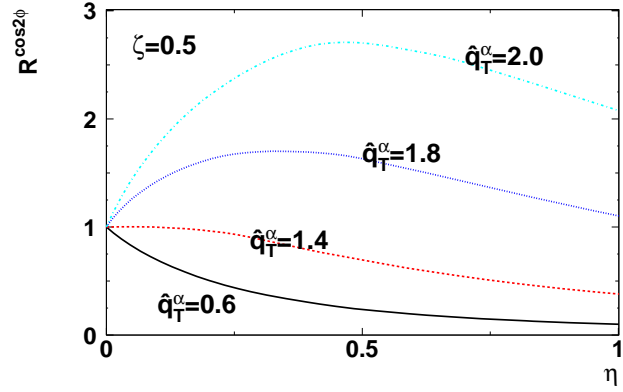}}
\caption{(Color online)  $R^{\cos 2\phi}$ {in Eq.(\ref{Rcos2phi})} as a
function of $\eta$ for different $\hat q_T^\alpha$ and the relative width $\zeta$  of the TMD quark distributions.  }
\label{fig:AfH}       
\end{figure}

The nuclear effect of the azimuthal asymmetry $\cos 2\phi$ can be measured by the ratio,
\begin{eqnarray}
\label{Rcos2phi}
R^{\cos 2\phi}\equiv \frac{A_{NA}^{\cos2\phi}}{A_{NN}^{\cos2\phi}}=
\frac{2\alpha+\Delta_{2F}}{2\alpha}
\left(\frac{2\beta}{2\beta+\Delta_{2F}}\right)^3
e^{\frac{(\alpha-\beta)
(2\alpha+2\beta+\Delta_{2F})\Delta_{2F}}
{2\alpha\beta
(2\alpha+\Delta_{2F})(2\beta+\Delta_{2F})}\vec{q}_{T}^{\ 2}}.
\end{eqnarray}
In the special case where  { $\alpha=\beta$}, we can obtain a simplified result
\begin{eqnarray}
\label{specialazi}
R^{\cos 2\phi}&=&
\left(\frac{2\alpha}{2\alpha+\Delta_{2F}}\right)^2,
\end{eqnarray}
which means that the azimuthal asymmetry $\cos 2\phi$ in the DY process in nucleon-nucleus
collisions is suppressed compared to that in nucleon-nucleon collisions and has no dependence
on the transverse momentum of the lepton pairs. It is very interesting that
the suppression in Eq. (\ref{specialazi}) is  very similar way to  that of azimuthal asymmetry
$\cos 2\phi$ in SIDIS obtained in Ref. \cite{Song:2010pf}. For the general case, the nuclear modification
factor can only depend on three independent variables:
\begin{eqnarray}
\eta\equiv {\Delta_{2F}}/2{\alpha},\ \ \
\hat q_T^\alpha\equiv {|\vec q_T|}/\sqrt{2\alpha},\ \ \  \zeta\equiv {\beta}/{\alpha}.
\end{eqnarray}
The numerical results are plotted in Fig.~(\ref{fig:AfH}),
with $\zeta=2$ in panel (a) and 0.5 in panel (b), respectively, as functions of $\eta$,
at different  scaled transverse momenta $\hat q_T^\alpha$. We can see that they are very similar to the results that have been obtained in
Refs. \cite{Gao:2010mj,Song:2010pf}. In the case $\zeta>1$, the azimuthal asymmetry is suppressed
and the suppression increases with the transverse momentum $\hat q_{T}^\alpha$. When $\zeta<1$, the suppression
will have the opposite dependence on  $\hat q_{T}^\alpha$.
Especially, the azimuthal asymmetry could be enhanced, instead of suppression,
for large enough transverse momentum $\hat q_{T}^\alpha$.
It means that the nuclear modification of the azimuthal asymmetry and its transverse
momentum dependence provides  a very sensitive probe to measure
the width of the transverse momentum distribution in
the TMD quark distribution functions.
{A numerical estimate of the magnitude of $\Delta_{2F}$ has been made very recently \cite{Song:2014sja}
by using the empirical result for the transport parameter $\hat q$ obtained from jet quenching in cold nuclei.
Here, one uses $\Delta_{2F}\approx \hat{q} L_A$ and
$\hat{q} \approx 0.025 \textrm{GeV}^2/\textrm{fm}$  \cite{Wang:2009qb,Chang:2014fba},
and the average of the transport distance $L_A$ in a nucleus is $L_A=3R_A/4=3 R_N A^{1/3}/4$.
Take Au as an example, one obtains $\Delta_{2F}\sim 0.12\textrm{GeV}^2$,
which leads to $\eta\sim0.24$ if we choose $\alpha=0.25\textrm{GeV}^2$. }

We can also calculate the {$q_T$-integrated}  azimuthal asymmetry,
which is given by
\begin{eqnarray}
\bar A_{NA}^{\cos2\phi}=
 \frac{\int d^2\vec q_T  \mathcal{F}_1\left[ h_{1}^{\perp}, h_{1}^{\perp}\right]}
   {\int d^2\vec q_T 4\mathcal{F}_0\left[f_{1},f_{1}\right]}=
\frac{1}{4M^2}\frac{\beta^2}{(2\beta+\Delta_{2F})}
\frac{\mathcal{S}\left[h_{1}^{\perp}{h}_{1}^{\perp}\right]}{\mathcal{S}\left[f_{1}{f}_{1}\right]}.
\end{eqnarray}
Therefore the {$q_T$-integrated}  nuclear modification factor of the azimuthal asymmetry $\cos 2\phi$ reads
\begin{eqnarray}
\label{average1}
 \bar R^{\cos 2\phi}\equiv \frac{\bar A_{NA}^{\cos2\phi}}{\bar A_{NN}^{\cos2\phi}} &=& \frac{2\beta}{2\beta+\Delta_{2F}}.
\end{eqnarray}

Now let us turn to the azimuthal asymmetries associated with the polarization of the incident nucleon, we can see
four more distribution functions are involved in
 Eq. (\ref{sigma}): $h_{1L}^{\perp}$, $f_{1T}^{\perp}$, $h_{1}$ and  $h_{1T}^{\perp}$. Once more, to illustrate
 the nuclear dependence of all these azimuthal asymmetries resulting from  these TMD distributions, we make  four more Gaussian ansatz assumptions
\begin{eqnarray}
 h_{1L}^{N\perp}(x,\vec{k}_{T})&=&\frac{1}{\pi \sigma_1}h_{1L}^{N\perp}(x)
e^{\frac{-\vec k_{T}^2}{\sigma_1}},\ \ \
f_{1T}^{N\perp}(x,\vec{k}_{T})=\frac{1}{\pi \sigma_2}f_{1T}^{N\perp}(x)
e^{\frac{-\vec k_{T}^2}{\sigma_2}},\\
h_{1}^{N}(x,\vec{k}_{T})&=&\frac{1}{\pi \sigma_3}h_{1}^{N}(x)
e^{\frac{-\vec k_{T}^2}{\sigma_3}},\ \ \
h_{1T}^{N\perp}(x,\vec{k}_{T})=\frac{1}{\pi \sigma_4}
h_{1T}^{N\perp}(x)e^{\frac{-\vec k_{T}^2}{\sigma_4}}.
\end{eqnarray}
Following the same routine, we can have
\begin{eqnarray}
A_{NA}^{\sin 2\phi}&=&\frac{\lambda_s}{4M^2}\frac{\sigma_1 {\beta}(2\alpha+\Delta_{2F})}
{(\sigma_1+{\beta}+\Delta_{2F})^3}
\frac{\mathcal{S}\left[h_{1L}^{\perp},{h}_{1}^{\perp}\right]}{\mathcal{S}\left[f_{1},{f}_{1}\right]}\vec{q}_{T}^{\ 2}
e^{\frac{\sigma_1+\beta-2\alpha}{(2\alpha+\Delta_{2F})
(\sigma_1+{\beta}+\Delta_{2F})}\vec{q}_{T}^{\ 2}},\\
A_{NA}^{\sin \phi_s}&=&-\frac{|\vec s_T|}{2M}\frac{\sigma_2(2\alpha+ \Delta_{2F})}
{(\sigma_2+{\alpha}+ \Delta_{2F})^2}
\frac{\mathcal{S}\left[f_{1T}^{\perp}, f_{1}\right]}{\mathcal{S}\left[f_{1},{f}_{1}\right]}|\vec{q}_{T}|
e^{\frac{\sigma_2-\alpha}{(2\alpha+\Delta_{2F})
(\sigma_2+{\alpha}+\Delta_{2F})}\vec{q}_{T}^{\ 2}},\\
A_{NA}^{\sin (2\phi-\phi_s)}&=&\frac{|\vec s_T|}{4M}\frac{\beta(2\alpha+ \Delta_{2F})}{(\sigma_3+{\beta}+ \Delta_{2F})^2}
\frac{\mathcal{S}\left[h_{1}, h_{1}^{\perp}\right]}{\mathcal{S}\left[f_{1},{f}_{1}\right]}
|\vec{q}_{T}|e^{\frac{\sigma_3+\beta-2\alpha}{(2\alpha+\Delta_{2F})
(\sigma_3+{\beta}+\Delta_{2F})}\vec{q}_{T}^{\ 2}},\\
A_{NA}^{\sin (2\phi+\phi_s)}&=&\frac{|\vec s_T|}{8M^3}
\frac{\sigma_4^2\beta (2\alpha+\Delta_{2F})}{(\sigma_4+\beta+\Delta_{2F})^4}
\frac{\mathcal{S}\left[h_{1T}^{\perp},h_{1}^{\perp}\right]}{\mathcal{S}\left[f_{1},{f}_{1}\right]}|\vec{q}_{T}|^3
e^{\frac{\sigma_4+\beta-2\alpha}{(2\alpha+\Delta_{2F})
(\sigma_4+{\beta}+\Delta_{2F})}\vec{q}_{T}^{\ 2}}.\hspace{0.9cm}
\end{eqnarray}
The nuclear modification factors corresponding to the above  different azimuthal asymmetries are given by, respectively,
\begin{eqnarray}
R^{\sin2\phi}&\equiv&\frac{A_{NA}^{\sin 2\phi}}{A_{NN}^{\sin 2\phi}}=
\frac{(2\alpha+\Delta_{2F})(\sigma_1+\beta)^3}{2\alpha(\sigma_1+\beta+\Delta_{2F})^3}
e^{-\frac{(\sigma_1+\beta-2\alpha)(\sigma_1+\beta+2\alpha+\Delta_{2F})\Delta_{2F}}
{2\alpha(2\alpha+\Delta_{2F})(\sigma_1+\beta)(\sigma_1+\beta+\Delta_{2F})}\vec q_T^{\ 2}},\\
R^{\sin\phi_s}&\equiv&\frac{A_{NA}^{\sin \phi_s}}{A_{NN}^{\sin \phi_s}}=
\frac{(2\alpha+\Delta_{2F})(\sigma_2+\alpha)^2}{2\alpha(\sigma_2+\alpha+\Delta_{2F})^2}
e^{-\frac{(\sigma_2-\alpha)(\sigma_2+3\alpha+\Delta_{2F})\Delta_{2F}}
{2\alpha(2\alpha+\Delta_{2F})(\sigma_2+\alpha)(\sigma_2+\alpha+\Delta_{2F})}\vec q_T^{\ 2}},\\
R^{\sin(2\phi-\phi_s)}&\equiv&\frac{A_{NA}^{\sin (2\phi-\phi_s)}}{A_{NN}^{\sin (2\phi-\phi_s)}}=
\frac{(2\alpha+\Delta_{2F})(\sigma_3+\beta)^2}{2\alpha(\sigma_3+\beta+\Delta_{2F})^2}
e^{-\frac{(\sigma_3+\beta-2\alpha)(\sigma_3+\beta+2\alpha+\Delta_{2F})\Delta_{2F}}
{2\alpha(2\alpha+\Delta_{2F})(\sigma_3+\beta)(\sigma_3+\beta+\Delta_{2F})}\vec q_T^{\ 2}},\\
R^{\sin(2\phi+\phi_s)}&\equiv&\frac{A_{NA}^{\sin (2\phi+\phi_s)}}{A_{NN}^{\sin (2\phi+\phi_s)}}=
\frac{(2\alpha+\Delta_{2F})(\sigma_4+\beta)^4}{2\alpha(\sigma_4+\beta+\Delta_{2F})^4}
e^{-\frac{(\sigma_4+\beta-2\alpha)(\sigma_4+\beta+2\alpha+\Delta_{2F})\Delta_{2F}}
{2\alpha(2\alpha+\Delta_{2F})(\sigma_4+\beta)(\sigma_4+\beta+\Delta_{2F})}\vec q_T^{\ 2}}.\hspace{1.2cm}
\end{eqnarray}
In the special cases of $\sigma_1=\sigma_3=\sigma_4=2\alpha-\beta$ and $\sigma_2=\alpha$, we can reduce them to
\begin{eqnarray}
R^{\sin2\phi}&=&
\left(\frac{2\alpha}{2\alpha+\Delta_{2F}}\right)^2,\hspace{1.2cm}
R^{\sin\phi_s}=
\frac{2\alpha}{2\alpha+\Delta_{2F}},\\
R^{\sin(2\phi-\phi_s)}&=&
\frac{2\alpha}{2\alpha+\Delta_{2F}},\hspace{1.3cm}
R^{\sin(2\phi+\phi_s)}=
\left(\frac{2\alpha}{2\alpha+\Delta_{2F}}\right)^3.
\end{eqnarray}
For the general cases, we can choose three independent variables  for every nuclear modification factor
like we did for the azimuthal asymmetry $\cos 2\phi$. We  choose two same scaled variables
$\eta\equiv\Delta_{2F}/2\alpha$ and $\hat q^\alpha_T \equiv |\vec q_T|/\sqrt{2\alpha}$ for
all the nuclear modification factors, and the rest,  $\zeta\equiv (\sigma_1+\beta)/2\alpha$, $\sigma_2/\alpha$, $(\sigma_3+\beta)/2\alpha$ and
$(\sigma_4+\beta)/2\alpha$, correspond to the $\sin2\phi$, $\sin\phi_s$, $\sin(2\phi-\phi_s)$, and $\sin(2\phi+\phi_s)$, respectively.
With these variables, it is obvious that
$R^{\cos2\phi}$ and $R^{\sin2\phi}$ are the same functions while $R^{\sin\phi_s}$
and  $R^{\sin(2\phi-\phi_s)}$ are the same. It should be noted that $R^{\sin\phi_s}$
and  $R^{\sin(2\phi-\phi_s)}$
 are very similar to the result of the azimuthal asymmetry $\cos \phi$ in SIDIS obtained in Ref.\cite{Gao:2010mj} and
$R^{\cos2\phi}$ and $R^{\sin2\phi}$  are very similar to the result of the azimuthal asymmetry $\cos 2\phi$ in SIDIS obtained in Ref. \cite{Song:2010pf}.
Besides, in the new scaled variables, the only difference between different azimuthal asymmetries  is up to an overall factor $\zeta/(\zeta+\eta)$
with right power order. Because our calculation
is only qualitative,  we do not show their numerical results in plots one by one. The shapes and features of these
azimuthal asymmetries associated with the polarization  are very similar to the unpolarized azimuthal asymmetry $\cos 2\phi$ shown in
Fig. \ref{fig:AfH}.
The {$q_T$-integrated} angular asymmetries  can be obtained in a very straightforward way:
\begin{eqnarray}
\bar A_{NA}^{\sin 2\phi}&=&
 \lambda_s \frac{\int d^2 \vec q_T \mathcal{F}_1\left[  h_{1L}^{\perp}, h_{1}^{\perp}\right]} {4\int d^2 \vec q_T\mathcal{F}_0\left[f_{1},f_{1}\right]}
=\frac{\lambda_s}{4M^2}\frac{\sigma_1 {\beta}}
{(\sigma_1+{\beta}+\Delta_{2F})}
\frac{\mathcal{S}\left[h_{1L}^{\perp},{h}_{1}^{\perp}\right]}{\mathcal{S}\left[f_{1},{f}_{1}\right]},\\
\bar A_{NA}^{\sin \phi_s}&=& -|\vec{s}_{T}|\frac{\int d^2 \vec q_T \mathcal{F}_2\left[ f_{1T}^{\perp},f_{1}\right]}
{2\int d^2 \vec q_T \mathcal{F}_0\left[f_{1}, f_{1}\right]}
=-\frac{|\vec s_T|}{2M}\frac{\sqrt{\pi}\sigma_2}
{2\sqrt{(\sigma_2+{\alpha}+ \Delta_{2F})}}
\frac{\mathcal{S}\left[f_{1T}^{\perp},f_{1}\right]}{\mathcal{S}\left[f_{1},{f}_{1}\right]},\\
\bar A_{NA}^{\sin (2\phi-\phi_s)}&=&|\vec{s}_{T}|\frac{\int d^2 \vec q_T \mathcal{F}_3\left[h_{1}, h_{1}^{\perp}\right]}
 {4\int d^2 \vec q_T\mathcal{F}_0\left[f_{1}, f_{1}\right]}
=\frac{|\vec s_T|}{4M}\frac{\sqrt{\pi}\beta}{2\sqrt{(\sigma_3+{\beta}+ \Delta_{2F})}}
\frac{\mathcal{S}\left[h_{1},h_{1}^{\perp}\right]}{\mathcal{S}\left[f_{1},{f}_{1}\right]},\\
\bar A_{NA}^{\sin (2\phi+\phi_s)}&=&
|\vec{s}_{T}|\frac{\int d^2 \vec q_T\mathcal{F}_4\left[ h_{1T}^{\perp}, h_{1}^{\perp}\right]}
{4\int d^2 \vec q_T\mathcal{F}_0\left[f_{1},f_{1}\right]}
=\frac{|\vec s_T|}{8M^3}
\frac{3\sqrt{\pi}\sigma_4^2\beta }{4\sqrt{(\sigma_4+\beta+\Delta_{2F})^3}}
\frac{\mathcal{S}\left[h_{1T}^{\perp}, h_{1}^{\perp}\right]}{\mathcal{S}\left[f_{1},{f}_{1}\right]}.\hspace{1.2cm}
\end{eqnarray}
It follows that the {$q_T$-integrated} nuclear modifications of the single-spin azimuthal asymmetries read
\begin{eqnarray}
\bar R^{\sin2\phi}&=&
\frac{\sigma_1+\beta}{\sigma_1+\beta+\Delta_{2F}},\hspace{1.5cm}
\bar R^{\sin\phi_s}=
\left(\frac{\sigma_2+\alpha}{\sigma_2+\alpha+\Delta_{2F}}\right)^{1/2},\\
\bar R^{\sin(2\phi-\phi_s)}&=&
\left(\frac{\sigma_3+\beta}{\sigma_3+\beta+\Delta_{2F}}\right)^{1/2},\hspace{0.3cm}
\bar R^{\sin(2\phi+\phi_s)} =
\left(\frac{\sigma_4+\beta}{\sigma_4+\beta+\Delta_{2F}}\right)^{3/2}.\hspace{0.5cm}
\end{eqnarray}

{We would like to mention that, although there are not many data available,
there are indeed measurements carried out on azimuthal asymmetries in DY processes with nuclear targets.
There are in particular measurements on $\langle\cos2\phi\rangle$ in $\pi d$ and $\pi A$ collisions  \cite{Badier:1981zpc,Conway:1989prd,Guanziroli:1988zpc,Falciano:1986zpc}.
It is obvious that the results for the differential cross section given by Eq. (\ref{sigma})
are applicable to reactions with pion beams as long as we drop all the polarization-dependent terms
because a pion is a spin zero object.
We expect  formulas for $\langle\cos2\phi\rangle$ similar to those given by Eqs. (\ref{ANNcos2phi}) and (\ref{ANAcos2phi}) and also a
nuclear suppression effect similar to that given by Eq. (\ref{Rcos2phi}).
It is therefore clear that precise measurements of the asymmetry $\langle\cos2\phi\rangle$ in these collisions
will be useful not only in parameterizing Boer-Mulders function but also in studying the nuclear dependence.
The accuracy of the data\cite{Badier:1981zpc,Conway:1989prd,Guanziroli:1988zpc,Falciano:1986zpc}
available is however still too low to draw any decisive judgment on the $A$ dependence of the asymmetry.
Nevertheless, it is encouraging to see that such measurements can indeed be carried out
and more measurements are planned \cite{Chang:2013opa}.}

\section{summary}

Within the framework of the TMD factorization, the nuclear dependence of the azimuthal asymmetry in polarized DY processes has been studied.
We find the nuclear modifications of the azimuthal asymmetries $\cos 2\phi$ and $\sin 2\phi$ are the same and
very similar to the azimuthal asymmetries $\cos 2\phi$ in SIDIS obtained in Ref. \cite{Song:2010pf}.
 The nuclear effects of the azimuthal asymmetries $\sin\phi_s$ and $\sin (2\phi-\phi_s)$
are the same and similar to the azimuthal asymmetry $\cos \phi$ in SIDIS obtained in Ref. \cite{Gao:2010mj}.
Among all the azimuthal asymmetries we have considered, the nuclear dependence of the azimuthal asymmetry $\sin (2\phi+\phi_s)$
is most suppressed. The nuclear modification of the azimuthal asymmetry and its nontrivial transverse
momentum dependence  provides  a very sensitive probe to measure  the width of the transverse momentum distribution in
the TMD quark distribution functions.

\begin{acknowledgments}
This work is supported partially by the Major State
Basic Research Development Program in China (Grant No. 2014CB845406).
J.H.G. was supported in part by the National Natural Science Foundation of China
under  Grant No.~11105137 and the  CCNU-QLPL Innovation Fund (Grant NO. QLPL2011P01 and QLPL2014P01). L.C. and Z.T.L. were supported in part by the National Natural Science Foundation of China under  Grant No.~11035003.
\end{acknowledgments}


\begin{thebibliography}{00}


\bibitem{Georgi:1977tv}
  H.~Georgi and H.~Politzer,
  Phys.\ Rev.\ Lett.\  {\bf 40}, 3 (1978).


\bibitem{Cahn:1978se}
  R.~N.~Cahn,
  Phys.\ Lett.\ B {\bf 78}, 269 (1978).

\bibitem{Berger:1979kz}
 E.~L.~Berger,
 Phys.\ Lett.\ B {\bf 89}, 241 (1980).


\bibitem{Oganesian:1997jq}
 K.~A.~Oganesian, H.~R.~Avakian, N.~Bianchi and P.~Di Nezza,
  Eur.\ Phys.\ J.\ C {\bf 5}, 681 (1998).

\bibitem{Chay:1997qy}
  J.~Chay and S.~M.~Kim,
  Phys.\ Rev.\ D {\bf 57}, 224 (1998)




\bibitem{Collins:1977iv}
  J.~C.~Collins and D.~E.~Soper,
  Phys.\ Rev.\ D {\bf 16}, 2219 (1977).

\bibitem{Collins:1978yt}
  J.~C.~Collins,
  Phys.\ Rev.\ Lett.\  {\bf 42}, 291 (1979).



\bibitem{Lam:1978pu}
  C.~S.~Lam and W.~K.~Tung,
  Phys.\ Rev.\  D {\bf 18}, 2447 (1978).


\bibitem{Lam:1978zr}
  C.~S.~Lam and W.~-K.~Tung,
  Phys.\ Lett.\ B {\bf 80}, 228 (1979).

\bibitem{Lam:1980uc}
  C.~S.~Lam and W.~-K.~Tung,
  Phys.\ Rev.\ D {\bf 21}, 2712 (1980).

\bibitem{Fries:1999jj}
  R.~J.~Fries, B.~Muller, A.~Schafer, E.~Stein,
  Phys.\ Rev.\ Lett.\  {\bf 83}, 4261-4264 (1999).
  R.~J.~Fries, A.~Schafer, E.~Stein, B.~Muller,
  Nucl.\ Phys.\  {\bf B582}, 537-570 (2000).

\bibitem{Gelis:2006hy}
  F.~Gelis, J.~Jalilian-Marian,
  Phys.\ Rev.\  {\bf D76}, 074015 (2007).



\bibitem{Liang:2006wp}
  Z.~T.~Liang and X.~N.~Wang,
  Phys.\ Rev.\  D {\bf 75}, 094002 (2007)


\bibitem{Zhou:2009jm}
  J.~Zhou, F.~Yuan and Z.~-T.~Liang,
  Phys.\ Rev.\ D {\bf 81}, 054008 (2010)

\bibitem{Baier:1996sk}
  R.~Baier, Y.~L.~Dokshitzer, A.~H.~Mueller, S.~Peigne and D.~Schiff,
  Nucl.\ Phys.\  B {\bf 484}, 265 (1997).



  \bibitem{Bodwin:1988fs}
  G.~T.~Bodwin, S.~J.~Brodsky and G.~P.~Lepage,
  Phys.\ Rev.\  D {\bf 39}, 3287 (1989).

\bibitem{Luo:1992fz}
  M.~Luo, J.~-w.~Qiu, and G.~F.~Sterman,
  Phys.\ Lett.\  {\bf B279}, 377-383 (1992);
  Phys.\ Rev.\  {\bf D49}, 4493-4502 (1994);
  Phys.\ Rev.\  {\bf D50}, 1951-1971 (1994).

\bibitem{Guo:1998rd}
  X.~F.~Guo,
  Phys.\ Rev.\  D {\bf 58}, 114033 (1998).

\bibitem{Wiedemann:2000za}
  U.~A.~Wiedemann,
  Nucl.\ Phys.\  B {\bf 588}, 303 (2000)

\bibitem{Guo:2000nz}
  X.~-f.~Guo and X.~-N.~Wang,
  Phys.\ Rev.\ Lett.\  {\bf 85}, 3591 (2000)

\bibitem{Wang:2001ifa}
  X.~-N.~Wang and X.~-f.~Guo,
  Nucl.\ Phys.\ A {\bf 696}, 788 (2001)




\bibitem{Fries:2002mu}
  R.~J.~Fries,
  Phys.\ Rev.\  D {\bf 68}, 074013 (2003).

\bibitem{Majumder:2007hx}
  A.~Majumder, B.~Muller,
  Phys.\ Rev.\  {\bf C77}, 054903 (2008).

\bibitem{Liang:2008vz}
  Z.~T.~Liang, X.~N.~Wang and J.~Zhou,
  Phys.\ Rev.\  D {\bf 77}, 125010 (2008)



\bibitem{D'Eramo:2010ak}
  F.~D'Eramo, H.~Liu and K.~Rajagopal,
  Phys.\ Rev.\ D {\bf 84}, 065015 (2011)



\bibitem{D'Eramo:2011zz}
  F.~D'Eramo, H.~Liu, K.~Rajagopal,
  Nucl.\ Phys.\  {\bf A855}, 457-460 (2011).


\bibitem{D'Eramo:2011zzb}
  F.~D'Eramo, H.~Liu and K.~Rajagopal,
  J.\ Phys.\ G {\bf 38}, 124162 (2011).

\bibitem{Gao:2010mj}
  J.~-H.~Gao, Z.~-T.~Liang, X.~-N.~Wang,
  Phys.\ Rev.\  {\bf C81}, 065211 (2010).




\bibitem{Song:2010pf}
  Y.~-K.~Song, J.~-H.~Gao, Z.~-T.~Liang, X.~-N.~Wang,
  Phys.\ Rev.\  {\bf D83}, 054010 (2011).

\bibitem{Gao:2011mf}
  J.~-H.~Gao, A.~Schafer and J.~Zhou,
  Phys.\ Rev.\ D {\bf 85}, 074027 (2012).






\bibitem{Collins:1981uk}
  J.~C.~Collins and D.~E.~Soper,
  Nucl.\ Phys.\ B {\bf 193}, 381 (1981)
  [Erratum-ibid.\ B {\bf 213}, 545 (1983)]
  [Nucl.\ Phys.\ B {\bf 213}, 545 (1983)].


\bibitem{Ji:2004xq}
  X.~-d.~Ji, J.~-P.~Ma and F.~Yuan,
  Phys.\ Lett.\ B {\bf 597}, 299 (2004)


\bibitem{Boer:1999mm}
  D.~Boer,
  Phys.\ Rev.\ D {\bf 60}, 014012 (1999)

\bibitem{Arnold:2008kf}
  S.~Arnold, A.~Metz and M.~Schlegel,
  Phys.\ Rev.\ D {\bf 79}, 034005 (2009)
\bibitem{Lu:2011cw}
  Z.~Lu, B.~-Q.~Ma and J.~Zhu,
  Phys.\ Rev.\ D {\bf 84}, 074036 (2011)



\bibitem{Mulders:1995dh}
  P.~J.~Mulders and R.~D.~Tangerman,
  Nucl.\ Phys.\ B {\bf 461}, 197 (1996)
  [Erratum-ibid.\ B {\bf 484}, 538 (1997)]


\bibitem{Goeke:2005hb}
  K.~Goeke, A.~Metz and M.~Schlegel,
  Phys.\ Lett.\ B {\bf 618}, 90 (2005)

\bibitem{Bacchetta:2006tn}
  A.~Bacchetta, M.~Diehl, K.~Goeke, A.~Metz, P.~J.~Mulders and M.~Schlegel,
  JHEP {\bf 0702}, 093 (2007)

\bibitem{Sivers:1989cc}
  D.~W.~Sivers,
  Phys.\ Rev.\ D {\bf 41}, 83 (1990).

\bibitem{Sivers:1990fh}
  D.~W.~Sivers,
  Phys.\ Rev.\ D {\bf 43}, 261 (1991).

\bibitem{Boer:1997nt}
  D.~Boer and P.~J.~Mulders,
  Phys.\ Rev.\ D {\bf 57}, 5780 (1998)



\bibitem{Pasquini:2006iv}
  B.~Pasquini, M.~Pincetti and S.~Boffi,
  Phys.\ Rev.\  D {\bf 76}, 034020 (2007).


\bibitem{Anselmino:2007fs}
  M.~Anselmino, M.~Boglione, U.~D'Alesio, A.~Kotzinian, F.~Murgia, A.~Prokudin
  and C.~Turk,
  Phys.\ Rev.\  D {\bf 75}, 054032 (2007).

\bibitem{Anselmino:2008jk}
  M.~Anselmino, M.~Boglione, U.~D'Alesio, A.~Kotzinian, F.~Murgia, A.~Prokudin and S.~Melis,
  Nucl.\ Phys.\ Proc.\ Suppl.\  {\bf 191}, 98 (2009).

\bibitem{Gamberg:2003ey}
  L.~P.~Gamberg, G.~R.~Goldstein and K.~A.~Oganessyan,
  Phys.\ Rev.\  D {\bf 67}, 071504(R) (2003).
\bibitem{Gamberg:2007wm}
  L.~P.~Gamberg, G.~R.~Goldstein and M.~Schlegel,
  Phys.\ Rev.\ D\ {\bf 77}, 094016 (2008).

\bibitem{Bacchetta:2007wc}
A.~Bacchetta, L.P.~Gamberg, G.R.~Goldstein, A.~Mukherjee,
Phys. \ Lett. \ B \ {\bf 659}, 234 (2008).



\bibitem{Avakian:2007xa}
  H.~Avakian, S.~J.~Brodsky, A.~Deur and F.~Yuan,
  Phys. \ Rev. \ Lett. \ {\bf99}, 082001 (2007).



\bibitem{Pasquini:2008ax}
  B.~Pasquini, S.~Cazzaniga and S.~Boffi,
  Phys. Rev. D {\bf 78}, 034025 (2008).



\bibitem{Bacchetta:2008af}
  A.~Bacchetta, F.~Conti and M.~Radici,
  Phys.\ Rev.\  D {\bf 78}, 074010 (2008).

\bibitem{Courtoy:2008vi}
  A.~Courtoy, F.~Fratini, S.~Scopetta and V.~Vento,
  Phys.\ Rev.\  D {\bf 78} (2008) 034002.

\bibitem{Courtoy:2008dn}
  A.~Courtoy, S.~Scopetta and V.~Vento,
  Phys.\ Rev.\  D {\bf 79}, 074001 (2009).

\bibitem{Courtoy:2009pc}
  A.~Courtoy, S.~Scopetta and V.~Vento,
  Phys.\ Rev.\  D {\bf 80}, 074032 (2009).


\bibitem{Avakian:2008dz}
  H.~Avakian, A.~V.~Efremov, P.~Schweitzer and F.~Yuan,
  Phys. \ Rev.  \ D {\bf 78}, 114024 (2008).
\bibitem{Anselmino:2008sga}
  M.~Anselmino {\it et al.},
  Eur.\ Phys.\ J.\  A {\bf 39} (2009) 89;
M.~Anselmino, M.~Boglione, U.~D'Alesio, A.~Kotzinian,
F.~Murgia and A.~Prokudin,
  Phys.\ Rev.\  D {\bf 71}, 074006 (2005);
  M.~Anselmino, M.~Boglione, U.~D'Alesio, A.~Kotzinian,
F.~Murgia and A.~Prokudin,
  Phys.\ Rev.\  D {\bf 72}, 094007 (2005)
  [Erratum-ibid.\  D {\bf 72}, 099903 (2005)];



\bibitem{Arnold:2008ap}
  S.~Arnold, A.~V.~Efremov, K.~Goeke, M.~Schlegel and P.~Schweitzer,
  arXiv:0805.2137 [hep-ph].

\bibitem{Efremov:2009ze}
  A.~V.~Efremov, P.~Schweitzer, O.~V.~Teryaev and P.~Zavada,
  Phys.\ Rev.\  D {\bf 80}, 014021 (2009).



\bibitem{She:2009jq}
  J.~She, J.~Zhu and B.~Q.~Ma,
  Phys.\ Rev.\  D {\bf 79}, 054008 (2009).


\bibitem{Jakob:1997wg}
  R.~Jakob, P.~J.~Mulders and J.~Rodrigues,
  Nucl.\ Phys.\  A {\bf 626}, 937 (1997);

\bibitem{Efremov:2003eq}
  A.~V.~Efremov, K.~Goeke and P.~Schweitzer,
  Eur.\ Phys.\ J.\ C {\bf 32}, 337 (2003).



\bibitem{Yuan:2003wk}
  F.~Yuan,
  Phys.\ Lett.\  B {\bf 575}, 45 (2003).

\bibitem{Pobylitsa:2003ty}
  P.~V.~Pobylitsa,
  arXiv:hep-ph/0301236.

\bibitem{Efremov:2004qs}
  A.~V.~Efremov, K.~Goeke and P.~Schweitzer,
  Eur.\ Phys.\ J.\  C {\bf 35}, 207 (2004).


\bibitem{Pasquini:2010af}
  B.~Pasquini and F.~Yuan,
  Phys.\ Rev.\ D {\bf 81}, 114013 (2010)










\bibitem{Collins:1992kk}
  J.~C.~Collins,
  Nucl.\ Phys.\ B {\bf 396}, 161 (1993)


\bibitem{BLM93}
  Z.~-T.~Liang and T.~-C.~Meng,
  Z.\ Phys.\ A {\bf 344}, 171 (1992);
  C.~Boros, Z.~T.~Liang and T.~C.~Meng,
  Phys.\ Rev.\ Lett.\  {\bf 70}, 1751 (1993).

\bibitem{Brodsky:2002rv}
  S.~J.~Brodsky, D.~S.~Hwang and I.~Schmidt,
  Nucl.\ Phys.\ B {\bf 642}, 344 (2002)



\bibitem{Mkrtchyan:2007sr}
  H.~Mkrtchyan, P.~E.~Bosted, G.~S.~Adams, A.~Ahmidouch, T.~Angelescu, J.~Arrington, R.~Asaturyan and O.~K.~Baker {\it et al.},
  Phys.\ Lett.\ B {\bf 665}, 20 (2008)

\bibitem{Airapetian:2009ae}
  A.~Airapetian {\it et al.}  [HERMES Collaboration],
  Phys.\ Rev.\ Lett.\  {\bf 103}, 152002 (2009)


\bibitem{Avakian:2010ae}
  H.~Avakian {\it et al.}  [CLAS Collaboration],
  Phys.\ Rev.\ Lett.\  {\bf 105}, 262002 (2010)


\bibitem{Alekseev:2010rw}
  M.~G.~Alekseev {\it et al.}  [COMPASS Collaboration],
  Phys.\ Lett.\ B {\bf 692}, 240 (2010)










\bibitem{Collins:2005rq}
  J.~C.~Collins, A.~V.~Efremov, K.~Goeke, M.~Grosse Perdekamp, S.~Menzel, B.~Meredith, A.~Metz and P.~Schweitzer,
  Phys.\ Rev.\ D {\bf 73}, 094023 (2006)

\bibitem{Vogelsang:2005cs}
  W.~Vogelsang and F.~Yuan,
  Phys.\ Rev.\ D {\bf 72}, 054028 (2005)







\bibitem{Majumder:2007ne}
  A.~Majumder, R.~J.~Fries and B.~Muller,
  Phys.\ Rev.\ C {\bf 77}, 065209 (2008)

\bibitem{Schafer:2013mza}
  A.~Schafer and J.~Zhou,
  Phys.\ Rev.\ D {\bf 88}, 074012 (2013)


\bibitem{Song:2014sja}
  Y.~-k.~Song, Z.~-t.~Liang and X.~-N.~Wang,
  arXiv:1402.3042 [nucl-th].




\bibitem{Wang:2009qb}
  W.~-t.~Deng and X.~-N.~Wang,
  Phys.\ Rev.\ C {\bf 81}, 024902 (2010)
\bibitem{Chang:2014fba}
  N.~-B.~Chang, W.~-T.~Deng and X.~-N.~Wang,
  arXiv:1401.5109 [nucl-th].

\bibitem{Badier:1981zpc}
  J.~Badier {\it et al.}  [NA3 Collaboration],
  Z.\ Phys.\ C {\bf 11}, 195 (1981).



\bibitem{Conway:1989prd}
  J.~S.~Conway, C.~E.~Adolphsen, J.~P.~Alexander, K.~J.~Anderson, J.~G.~Heinrich, J.~E.~Pilcher, A.~Possoz and E.~I.~Rosenberg {\it et al.},
  Phys.\ Rev.\ D {\bf 39}, 92 (1989).

\bibitem{Guanziroli:1988zpc}
  M.~Guanziroli {\it et al.}  [NA10 Collaboration],
  Z.\ Phys.\ C {\bf 37}, 545 (1988).

\bibitem{Falciano:1986zpc}
  S.~Falciano {\it et al.}  [NA10 Collaboration],
  Z.\ Phys.\ C {\bf 31}, 513 (1986).

\bibitem{Chang:2013opa}
  See e.g., W.~-C.~Chang and D.~Dutta,
  Int.\ J.\ Mod.\ Phys.\ E {\bf 22}, 1330020 (2013), and references therein.

\end{thebibliography}
\end{document}